\documentclass{article} 
\usepackage{graphicx}        
\begin{document}
\centerline{\bf Influence of changing environment on }

\centerline {\bf Shennan-type evolution of stone-age cultural innovation}
\medskip

\noindent
F.W.S. Lima* and Dietrich Stauffer**

\medskip
\noindent
* Departamento de F\'{\i}sica,
Universidade Federal do Piau\'{\i},
64.049-550 Teresina, Piau\'{\i}, Brazil

\medskip
\noindent
** Laboratoire PMMH, \'Ecole Sup\'erieure de Physique et de Chimie
Industrielles, 10 rue Vauquelin, F-75231 Paris, France

\noindent
visiting from Institute for Theoretical Physics, Cologne University,
D-50923 K\"oln, Euroland
\medskip

{\small The computer simulations of Shennan (2001) are complemented by 
assuming the environment to change randomly. For moderate change rates,
fitness optimisation through evolution is still possible.}

\bigskip
Keywords: Evolution theory, stone age, cultural innovation, computer simulation

\medskip
Shennan (2001), based on a biological model of Peck et al (1997), showed that 
in large populations, evolution in a computer model can evolve to a higher
average fitness than in a smaller population. This influence of demography 
may be relevant for an easier spread cultural innovation in {\it homo
sapiens} $10^4 \dots 10^5$ years ago. Now we test whether this effect survives 
if the environment changes and if thus also the combination of optimal traits
giving the highest fitness changes continuously during the evolution.

Each individual $j$ in the Shennan model has $L$ traits $x_{ij}$ with real 
numbers $-\infty < x_{ij} < \infty$ and deviations $d_{ij} = {\rm abs}(x_{ij}
- e_i)$ from the optimal values $e_i$. The fitness or fertility is

$$ w_j = \exp[-\sum_{i=1}^L \; d_{ij}]  \quad .$$ 

We start the simulations with all $e_i = x_{ij}$ (Shennan (2001) set all $e_i$ 
to zero permanently). Then at each sweep through the population of $N$ 
individuals (constituting one time step or generation), each individual $j$ 
gives birth to one offspring (baby, pupil) with probability $w_j$, while
with probability $1-w_j$ instead the best-fitted individual produces one
offspring. Thereafter, all adults die, and the offspring become the new 
adults. The selection of the best-fitted, instead of any other, individual
avoids the extremely low and perhaps unrealistic fitnesses of Shennan's simple
model and follows the spirit of his oblique transmission of culture by teachers
instead of parents. Also, at each time step each individual has one randomly 
selected $x$-value changed by an amount taken randomly between $\pm 0.04$, 
which represents a useful or damaging innovation, simpler than Shennan (2001)
and Peck et al (1997).

The highest curve in our upper figure shows for $L=100$ the resulting fitness, 
averaged over all individuals and then over the second half of all time steps
(geometric mean over the population, arithmetic mean over time). As in Shennan
(2001), larger populations are seen to lead to larger overall fitness.

Now as a new aspect in the spirit of Cebrat et al (2009) we introduce a changing
environment by changing, with probability $R$ at each time step, also all 
optimal values $e_i, \; i=1, 2, \dots L$ by random amounts also inhomogeneously 
distributed between $\pm 0.08$.
For example, in the migration of {\it homo sapiens} the techniques to walk on
snow and ice became helpful only rather late, while dark skin became 
disadvantageous in the Arctic (see also Gibbons 2010).

Our figure shows reasonable reductions of the fitness for small $R$ while for 
$R > 0.01$ not much is left. Thus for survival, environmental change rates 
should be not higher than the innovation rate per individual, which is $1/L = 
0.01$ according to our above rule and our choice $L=100$. Indeed, in the lower
part of the figure for $L=10$, we got reasonable survival even for $R=1$. Fig.
2a shows more systematically the dependence on $L$ for larger $L$.

For sexual reproduction, as in Peck et al (1997), we divide the population 
into men and women; now the two fittest members of the population give birth
if they happen to be of different sex. Each of the $L$ traits of the child is
randomly selected from father or from mother, with probability 1/2 each. The
results are shown in Fig.2b.

Children learn not only from their parents (vertical transmission) but also 
from biologically unrelated adult teachers (oblique transmission). For
the latter case we replace the parent (asexual case) or one of the parents
(sexual case) by a randomly selected teacher from whom the child learns half
of the traits. Fig.3 shows this oblique transmission; for the sexual case
a surprising minimum near $L = 2500$ appears for some $R$; Fig.3c shows
the results for $R=0.001$ on a logarithmic scale.

We checked with numerous different random number seeds the gap at $L = 3500$ 
in this minimum of Fig.3c as a function of time. We found that for short times
the fitness equilibrates to $0.86 \pm 0.02$, and then jumps down to (nearly) 
zero. The jump time varies from 4 to several thousand. This behaviour is similar
to homogenous nucleation in metastable states, as for example in supersaturated
water vapour. The direction of this rapid transition is, however, opposite to 
the rapid improvement in human civilisation about 45,000 years ago (Owen et al
2009). (For other $L$ between 150 and 25,000 the results usually are more 
smooth).
\medskip

This simulation was triggered by the course of Prof. S. L. Kuhn at Cologne
University, winter term 2009/10, where the Shennan paper was read. We thank
Profs. Shennan, J. Richter and P.M.C. de Oliveira for  encouragement and 
suggestions, and CNPq for support of the Brazilian author. 

\bigskip
\centerline{\bf Appendix}

Two variants of the above asexual model have also been investigated:
Instead of the above one innovation per person at each iteration, one
may have one innovation for each of the $L$ traits of each person at each 
iteration (``per person'' or ``per trait''). And instead of the fitness being
exp($-D$) for one person, one may take it as exp($-D/L$), i.e. ``with'' instead
of ``without'' division. All four combinations are shown in the last figure:

    per trait, without: +

    per trait, with: x

    per person, with: stars

    per person, without: squares

\medskip
\noindent
the last one being the above standard case. These symbols refer to a population
of 1000 at an environmental change rate of 0.003.  Thus per trait instead of 
per person decreases the fitness, and with division by $L$ quite trivially
the fitness strongly increases. With a rate 0.3 and a ten times larger 
population, the two overlapping lines in the figure are obtained, showing 
that per person or per trait does not matter much for this higher change
rate.

If the division by $L$ is applied to the gap in Fig.3c for oblique sexual
transmission, then the fitness stays near 0.98 for $L$ between 1000 and 5000,
with no gap. 

Finally we let the population size fluctuate by a Verhulst factor, instead
of keeping it constant, for the standard case ``per person'' and ``without
division''. Thus for each of the 10,000 iterations we went again through
each individual and let it die with probability $N/K$ where $n$ is the 
current population and $K=1000$ is usually called the carrying capacity. If 
the individual survives this ``logistic'' danger, it produces one additional
offspring. We found that the actual populations fluctuate near $K/2$
and that about half of them die out before the 10,000 iterations are
finished, if $K$ is about 16 (and $L$ is 10 or 100, for $R$ between 0.001
and 1; $L= 1000$, 10,000 and 100,000 for $R=0.003$ only.)
Thus random fluctuations will hardly kill a population of more than
a dozen individuals. Note that we include also the dangers from the random
changes in the environment, but not catastrophes like volcano eruptions 
not described by our rather smooth environmental changes.

\begin{figure}[hbt]
\begin{center}
\includegraphics[angle=-90,scale=0.35]{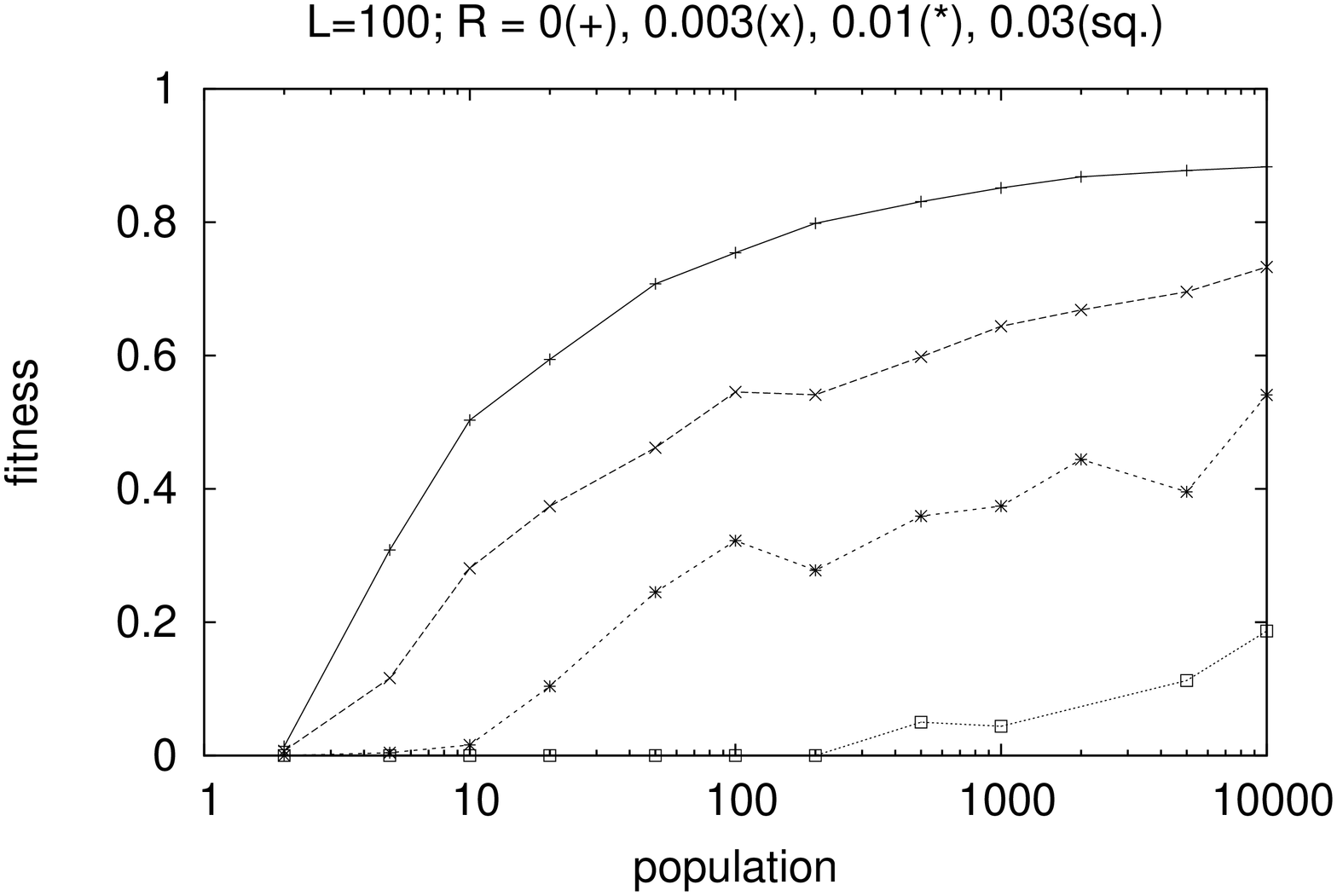}
\includegraphics[angle=-90,scale=0.35]{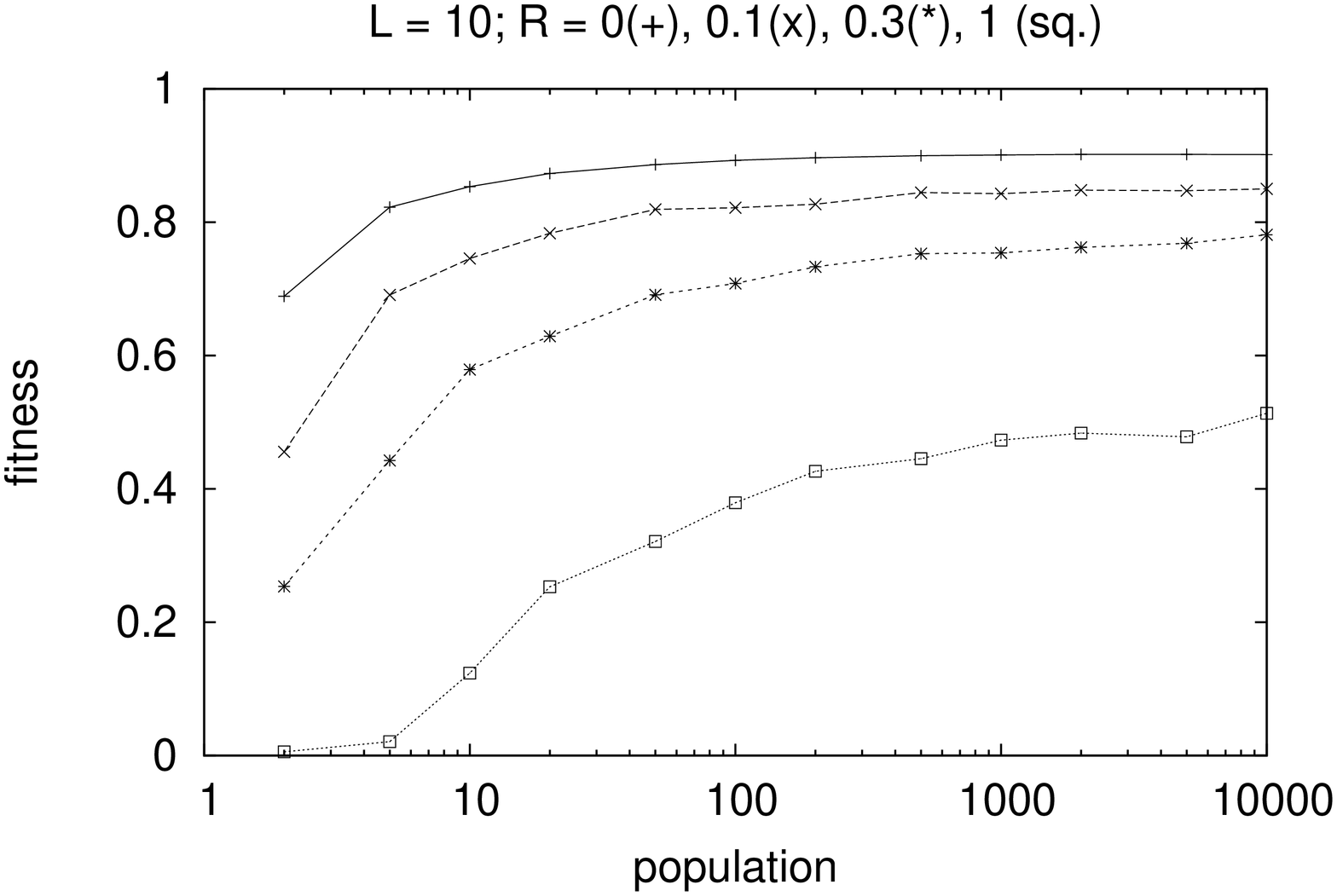}
\end{center}
\caption{Variation of average fitness with population size, for environmental
change rates $R$ increasing from top to bottom. ($10^4$ time steps.) Upper part
$L=100$, lower part $L=10$.
} 
\end{figure}

\begin{figure}[hbt]
\begin{center}
\includegraphics[angle=-90,scale=0.35]{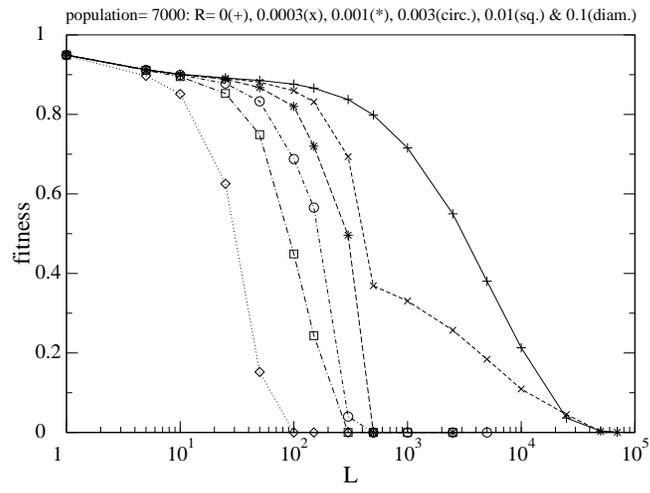}
\includegraphics[angle=-90,scale=0.35]{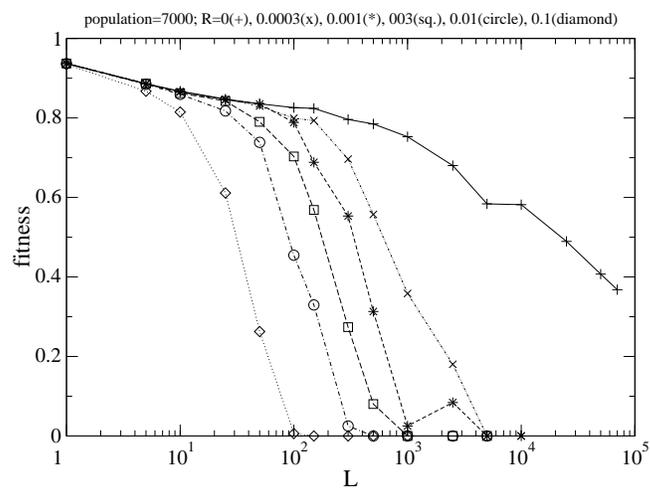}
\end{center}
\caption{Variation with the number $L$ of traits. Part a shows the asexual
and part b the sexual case. 
} 
\end{figure}

\begin{figure}[hbt]
\begin{center}
\includegraphics[angle=-90,scale=0.30]{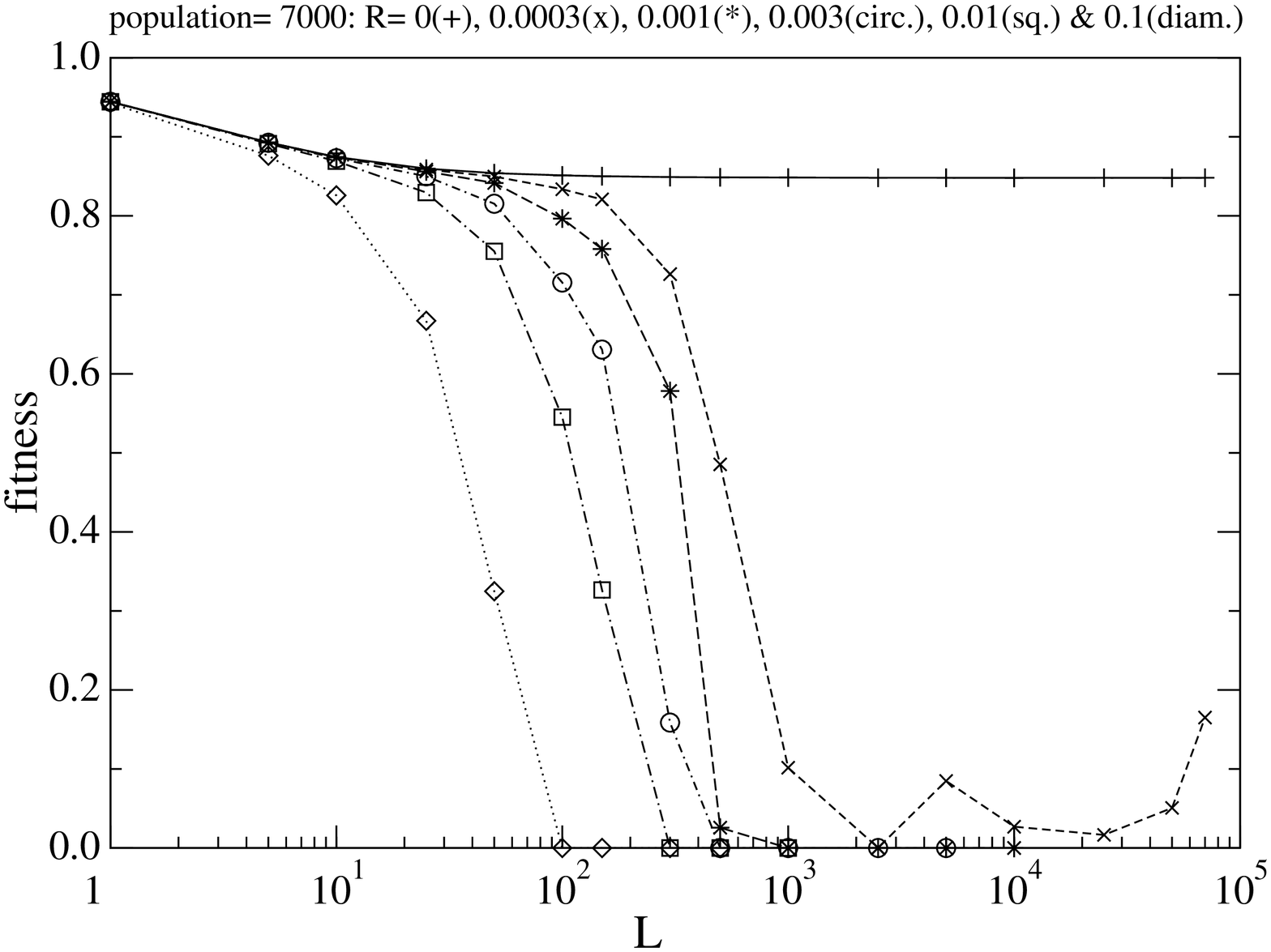}
\includegraphics[angle=-90,scale=0.30]{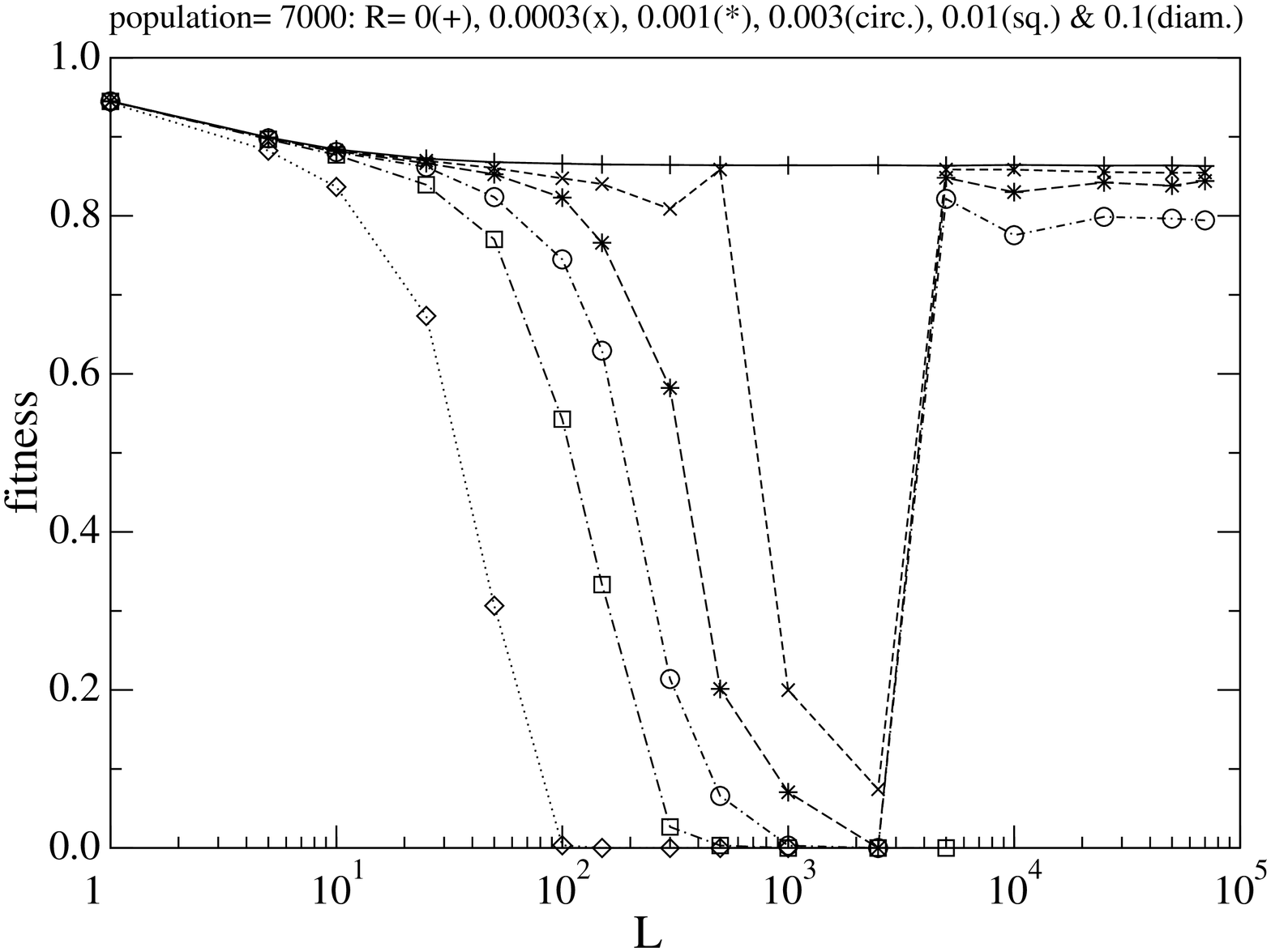}
\includegraphics[angle=-90,scale=0.30]{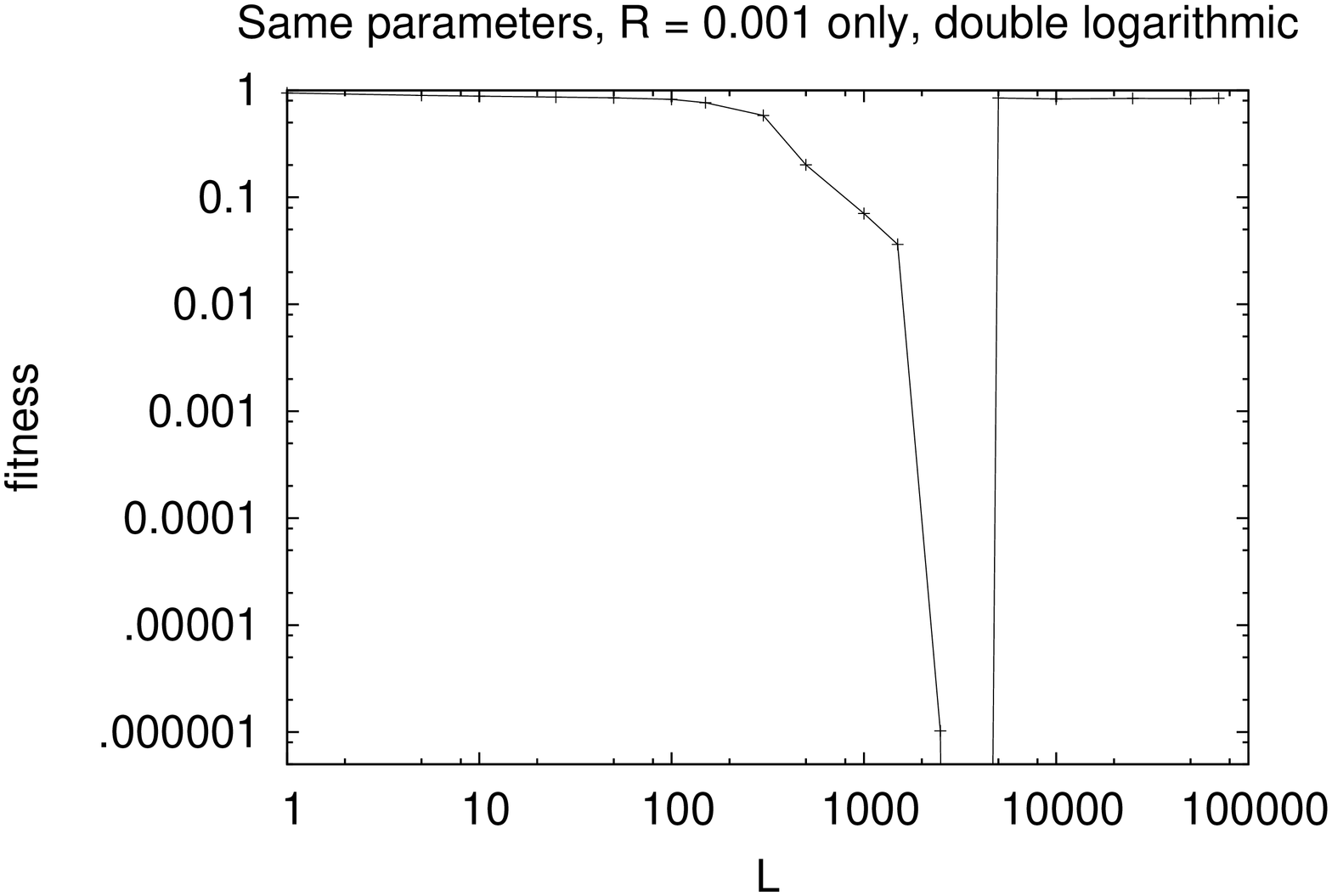}
\end{center}
\caption{ As Fig.2 but for oblique case: The child also learns from a teacher.  
The bottom part shows the fitness on a logarithmic scale for $R=0.001$ only.
} 
\end{figure}

\begin{figure}[hbt]
\begin{center}
\includegraphics[angle=-90,scale=0.5]{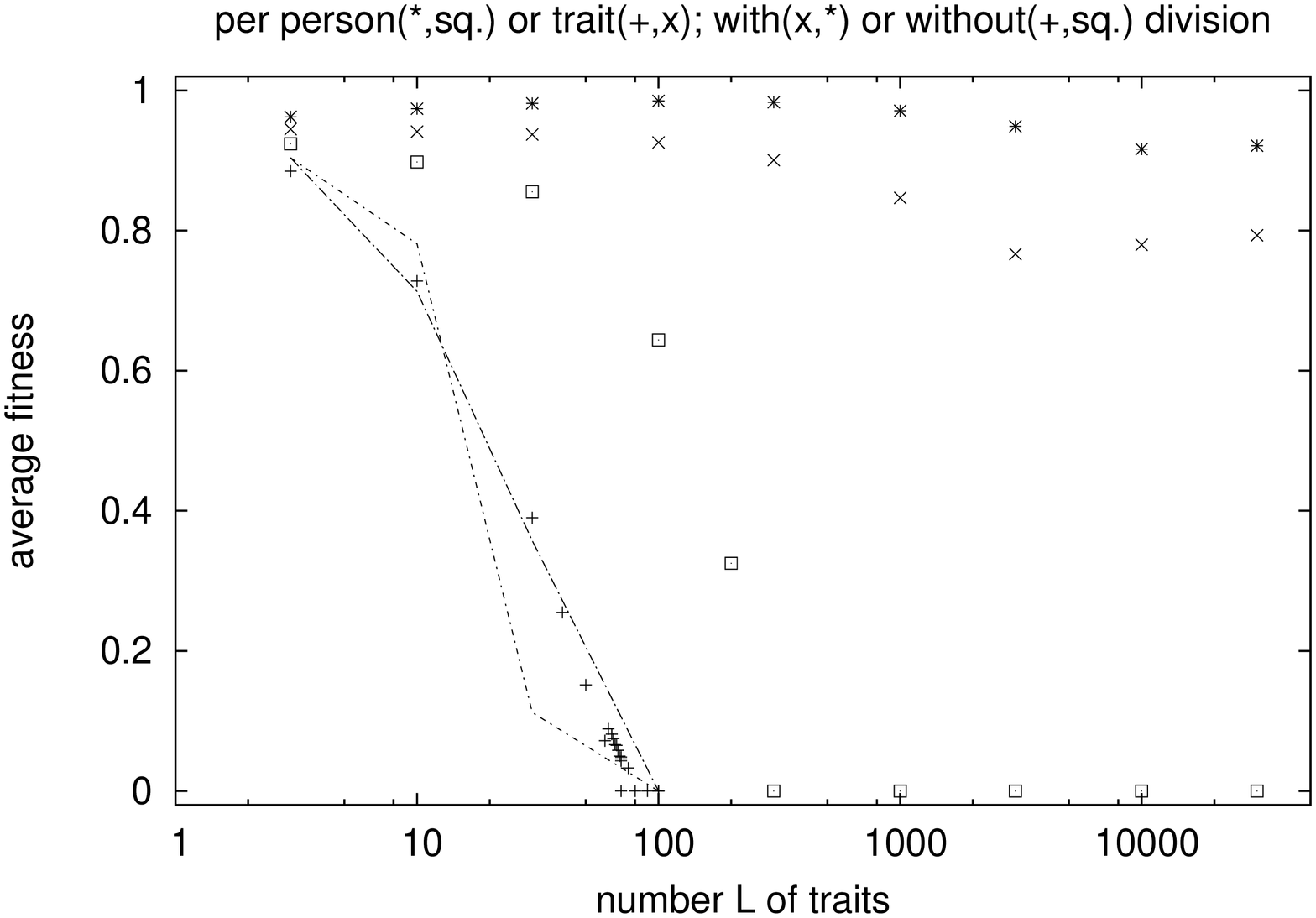}
\end{center}
\caption{As Fig.2a for the variants of our appendix
}
\end{figure}
\parindent 0pt
\bigskip

S. Cebrat, D. Stauffer, J.S. S\'a Martins, S. Moss de Oliveira and P.M.C.
de Oliveira, e-print arXiv:0911.0589 at arXiv.org (quantitative biology)
(2009). 

\medskip
A. Gibbons, Science 329, 740-742 (2010).

\medskip
J.R. Peck, G. Barreau and C.C. Heath, Genetics 145, 1171-1179 (1997).

\medskip
A. Powell, S. Shennan, and M. G. Thomas, Science 324, 1298-1301 (2009)

\medskip
S. Shennan, Cambridge Arch. J; 11, 5-16 (2001).


\end{document}